# TOPOLOGICAL FIELD THEORIES, STRING BACKGROUNDS AND HOMOTOPY ALGEBRAS

A. A. VORONOV*

*Department of Mathematics*
*University of Pennsylvania*
*Philadelphia, PA 19104-6395*
*USA**

December 22, 1993

**Abstract.** String backgrounds are described as purely geometric objects related to moduli spaces of Riemann surfaces, in the spirit of Segal's definition of a conformal field theory. Relations with conformal field theory, topological field theory and topological gravity are studied. For each field theory, an algebraic counterpart, the (homotopy) algebra satisfied by the tree level correlators, is constructed.

## 1. Introduction

The usual way of describing a string background as some construction on top of a conformal field theory involving the Virasoro operators, the antighost fields and the BRST operator appears too eclectic to be seriously accepted by the general mathematical public. Here we make an attempt to include string theory in the framework of geometric/topological field theories such as conformal field theory and topological field theory. Basically, we describe all two-dimensional field theories as variations on the theme of Segal's conformal field theories [9]. Our definition is in some sense dual to Segal's definition of a string background, also known as a topological conformal field theory, via differential forms and operator formalism, see Segal [10] and Getzler [1].

In this paper, each geometric field theory is followed by a leitmotif, the structure of an algebra built on the state space of the theory. Whereas it is commonly known that two-dimensional quantum field theories comprise very interesting geometrical structures, related algebraic structures have emerged

* Research supported in part by NSF grant DMS-9108269.A03
** Affiliated to Department of Mathematics, Princeton University, Princeton, NJ 08544-1000, USA

very recently and are still experiencing a very active period of growth. A list of references, perhaps, already outdated, can be found in the recent paper [5].

## 2. Topological Field Theory and Frobenius Algebras

*Note.* The field theories we are going to consider here will all have the total central charge zero. The general case can be done by involving the determinant line bundles over the moduli spaces.

A *topological field theory* (*TFT*) is a complex vector space $V$, called the *state space*, together with a correspondence

$$m\left\{\phantom{XXXXX}\right\}n \quad \longmapsto \quad |\Sigma\rangle : V^{\otimes m} \to V^{\otimes n} \tag{1}$$

An oriented surface $\Sigma$ bounding $m+n$ circles
A linear operator $|\Sigma\rangle$

Here a surface is not necessarily connected. Its boundary circles are enumerated and parameterized. The orientation of the first $m \geq 0$ circles, called *inputs*, is opposite to the orientation of $\Sigma$ and the orientation of the remaining $n \geq 0$ circles, called *outputs*, is compatible with the orientation induced from the surface. The linear operator $|\Sigma\rangle$ is called the *state* corresponding to the surface $\Sigma$.

This correspondence should satisfy the following axioms.

1. **Topological invariance**: The linear mapping $|\Sigma\rangle$ is invariant under orientation preserving diffeomorphisms of the surface $\Sigma$.
2. **Permutation equivariance**: The correspondence $\Sigma \mapsto |\Sigma\rangle$ commutes with the action of the symmetric groups $S_m$ and $S_n$ on surfaces and linear mappings by permutations of inputs and outputs.
3. **Factorization property**: Sewing along the parameterizations of the boundary corresponds to composing:

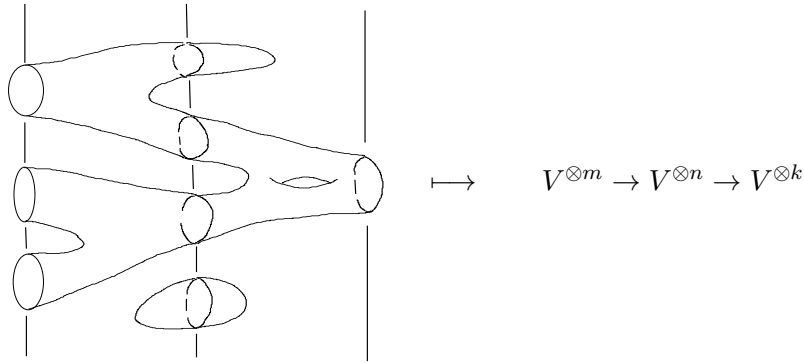

| The sewing of the outputs of a surface with inputs of another surface | The composition of the corresponding linear operators |

4. **Normalization**:

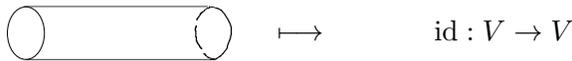

| A cylinder | The identity operator |

With the map $\mapsto$ id$: V \to V$

These data and axioms can be formulated equivalently using functors. Within this approach, a TFT is a multiplicative functor from a "topological" tensor category *Segal* to a "linear" tensor category *Hilbert*. An object of the category *Segal* is a diffeomorphism class of parameterized one-dimensional compact manifolds, i.e., disjoint unions of circles. A morphism between two collections of circles is a diffeomorphism class of oriented surfaces bounding the circles, so that the induced orientation is against the parameterizations on the first collection and compatible with that on the second collection of circles. The identity morphism of an object is the cylinder over it. The operation of disjoint union of collections of circles introduces the structure of a tensor category on *Segal*.

The other category *Hilbert* is the category of complex vector spaces (Hilbert in real examples), not necessarily finite dimensional, with the usual tensor product. Then the space $V$ is the vector space corresponding to the single circle and it is easily checked that the functoriality plays the role of the factorization property and that the two definitions are equivalent.

Any oriented surface can be cut into pants, caps and cylinders:

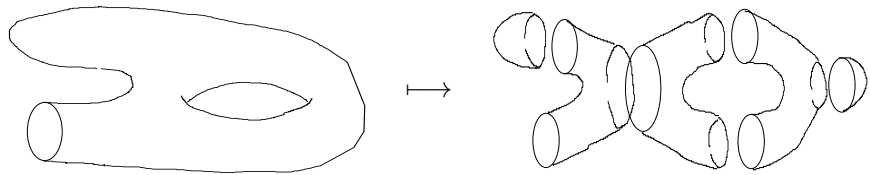

Thus, oriented surfaces have the following generators with respect the sewing operation. And respectively, the space $V$ is provided with an algebraic structure generated by the operations below with respect to composition of linear mappings.

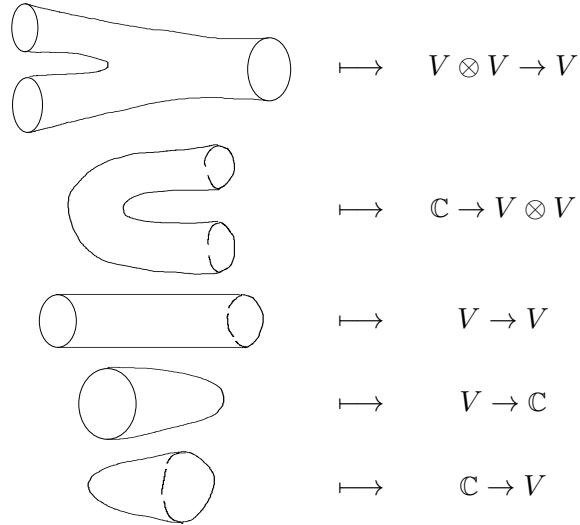

$$\begin{aligned} &\mapsto \quad V \otimes V \to V \\ &\mapsto \quad \mathbb{C} \to V \otimes V \\ &\mapsto \quad V \to V \\ &\mapsto \quad V \to \mathbb{C} \\ &\mapsto \quad \mathbb{C} \to V \end{aligned}$$

THEOREM 1 (Folklore). *A TFT is equivalent to a Frobenius algebra, i.e., a commutative algebra $V$ with a unity and a nondegenerate symmetric bilinear form $\langle,\rangle : V \otimes V \to \mathbb{C}$ which is invariant with respect to the multiplication:*

$$\langle ab, c \rangle = \langle a, bc \rangle$$

*and has an "adjoint" $\mathbb{C} \to V \otimes V$.*

An "adjoint" to a mapping $\phi : V \otimes V \to \mathbb{C}$ is a mapping $\psi : \mathbb{C} \to V \otimes V$, such that the compositions $V \xrightarrow{\mathrm{id} \otimes \psi} V \otimes V \otimes V \xrightarrow{\phi_{12} \otimes \mathrm{id}} V$ and $V \xrightarrow{\psi \otimes \mathrm{id}} V \otimes V \otimes V \xrightarrow{\mathrm{id} \otimes \phi_{23}} V$ are identities. When the space $V$ is finite dimensional, an inner product establishes an isomorphism $V \to V^*$, and an adjoint mapping gives a mapping $V^* \to V$, which is nothing but its inverse. Thus, in the finite dimensional case, a Frobenius algebra is just an algebra with an invaraint nondegenerate inner product. The theorem follows from the remark above about decomposing a surface into pants, caps and cylinders and the obvious fact that the symmetric form $\langle,\rangle$ in a Frobenius algebra $V$ can be obtained from a linear functional $f : V \to \mathbb{C}$ as $\langle a, b \rangle = f(ab)$.

An important substructure is observed for a TFT at the *tree level*, when we restrict our attention to surfaces of genus zero and with exactly one

output:

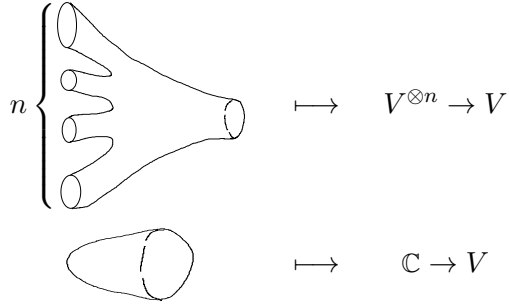

Topological invariance, permutation equivariance, the factorization and the normalization axioms make sense for such surfaces and are assumed.

The following fact is worth mentioning, because we are aiming to study similar algebraic structures occurring in string theory at the tree level.

COROLLARY 2. *A TFT at the tree level is equivalent to a commutative algebra $V$ with a unity.*

## 3. Conformal Field Theory

A *conformal field theory* (*CFT*) is a device very similar to a TFT, except that
1. the correspondence (1) is defined on Riemann surfaces bounding holomorphic disks and the state $|\Sigma\rangle$ depends smoothly on the Riemann surface $\Sigma$,
2. topological invariance is replaced by conformal invariance,
3. when two Riemann surfaces are sewn, the result is provided with a unique complex structure, and
4. normalization is slightly different:

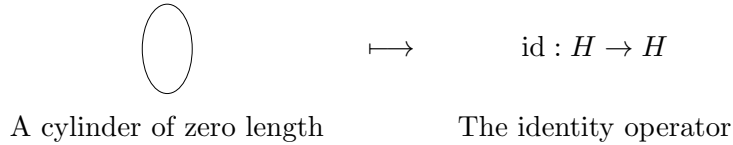

A cylinder of zero length        The identity operator

In other words, a CFT is a smooth mapping

$$\mathcal{P}_{m+n} \to \mathrm{Hom}(H^{\otimes m}, H^{\otimes n}), \quad (2)$$

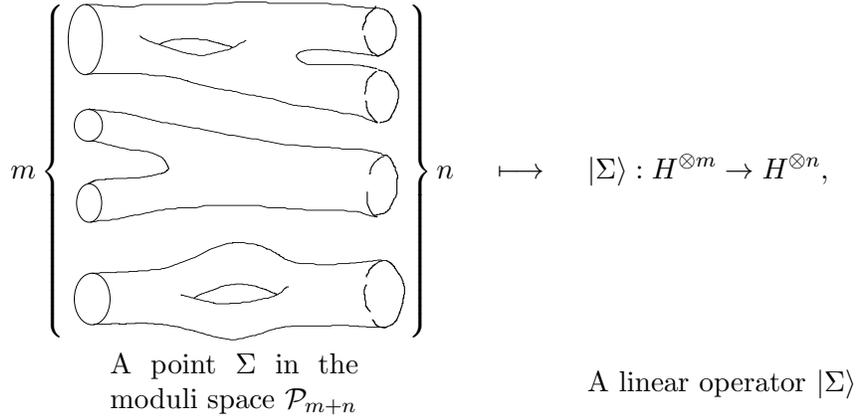

A point $\Sigma$ in the moduli space $\mathcal{P}_{m+n}$        A linear operator $|\Sigma\rangle$

$$|\Sigma\rangle : H^{\otimes m} \to H^{\otimes n},$$

where $\mathcal{P}_{m+n}$ is the moduli space of Riemann surfaces (one-dimensional complex compact manifolds) bounding $m$ negatively oriented holomorphic disks and $n$ positively oriented disks. The surfaces can have arbitrary genera, the disks are holomorphic mappings from the unit disk to a closed Riemann surface and they are enumerated. The mapping (2) must be equivariant with respect to permutations, transform sewing of Riemann surfaces into composition of the corresponding linear operators and must be normalized as above.

There is an evident reformulation of the CFT data as a functor from a suitable category *Segal* to the category *Hilbert* analogous to the one for TFT's.

## 4. String Theory and Homotopy Lie Algebras

### 4.1. STRING BACKGROUNDS

Let $H$ be a graded vector space with a differential $Q$, $Q^2 = 0$, i.e., $H$ be a complex. A *string background* is a correspondence

$$C_\bullet \mathcal{P}_{m+n} \to \mathrm{Hom}(H^{\otimes m}, H^{\otimes n}), \qquad (3)$$

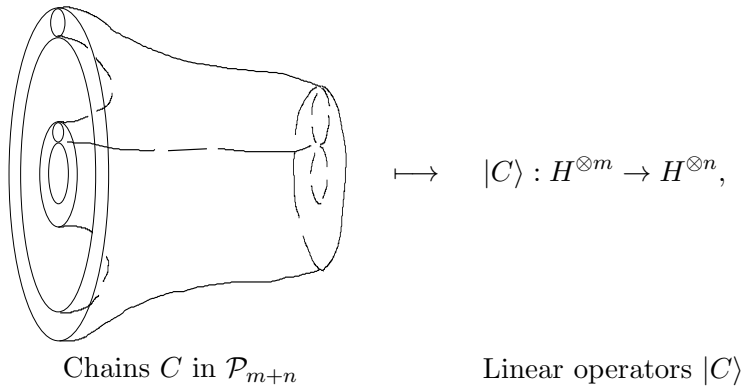

$$\mapsto \quad |C\rangle : H^{\otimes m} \to H^{\otimes n},$$

Chains $C$ in $\mathcal{P}_{m+n}$        Linear operators $|C\rangle$

which satisfies the axioms below. [On the figure, the surface is nothing but a pair of pants (so $m = 2$, $n = 1$) and the chain is just a circle. The pants moving along the circle in the moduli space sweep out a "surface of revolution", which I attempted to sketch above.] By chains here we mean the (complex) vector space generated by oriented (sectionally) smooth chains.

1. **Smoothness**: The mapping (3) is smooth.
2. **Equivariance**: The mapping (3) is equivariant with respect to permutations of inputs and outputs.
3. **Factorization**: The sewing of outputs of a chain with inputs of another chain (namely, outputs of each Riemann surface in the first chain are sewn with inputs of each Riemann surface in the second chain, each time producing a new Riemann surface) transforms under (3) into the composition of the corresponding linear operators.
4. **Homogeneity and $Q$-$\partial$-Invariance**: The mapping (3) is a morphism of complexes. That means that it maps a chain of dimension $k$ to a linear mapping of degree $-k$ (with respect to the natural grading on the Hom) and that the boundary of a chain in $\mathcal{P}_{m+n}$ transforms into the differential of the corresponding mapping,
$$|\partial C\rangle = Q|C\rangle,$$
where $Q$ acts on each of the $m+n$ components $H$ of Hom, as usual.
5. **Normalization**: The point {Riemann sphere with two unit disks around 0 and $\infty$ cut out} $\in \mathcal{P}_{m+n}$ maps to the identity operator $\mathrm{id}: H \to H$.

This correspondence can also be axiomatized as a functor, like in the cases of TFT and CFT. The corresponding category *Segal* will still have disjoint unions of circles as objects, but its morphisms will be chains in the moduli spaces. In the category *Hilbert*, one has to consider graded spaces with differentials (i.e., complexes), but still all linear mappings as morphisms.

The Virasoro semigroup of cylinders (including the group of diffeomorphisms of the circle, represented by cylinders of zero width) acts on $H$ via the degree 0 states $\exp tT(v) = |\exp tv\rangle$ corresponding to cylinders, regarded as points in $\mathcal{P}_{1+1}$:

$$\bigcirc\!\!\!\!\!\!\!\!\!\!\!\!\!\!\!\!\!\!\!\!\!\!\!\!\!\!\!\!\!\!\!\!\!\!\!\!\!\!\!\!\!\!\!\!\!\! \longmapsto \quad \exp tT(v): H \to H, \tag{4}$$
$$\exp tv$$

where $v$ is the generating complex vector field on the circle. The so-called *antighost* operators $b(v)$ on $H$ can also be easily identified in our picture. They are the derivatives
$$b(v) = \frac{d}{dt}B(tv)|_{t=0} \tag{5}$$
of the operators $B(tv)$ of degree $-1$ obtained when the same cylinder corresponding to $v$ is regarded as a one-chain in $\mathcal{P}_{1+1}$. At time $t$, the cylinder $\exp(tv)$ is a point in $\mathcal{P}_{1+1}$. When $t$ changes, these points sweep out a path in $\mathcal{P}_{1+1}$. Note that

$$[T(v_1), T(v_2)] = T([v_1, v_2]),$$

because the operators $\exp tT(v)$ define a representation of the Virasoro semigroup, and

$$\{b(v_1), b(v_2)\} = 0,$$

because the two-chains $\exp(sv_1) \times \exp(tv_2)$ and $\exp(tv_2) \times \exp(sv_1)$ differ only by orientation. In particular, $b^2(v) = 0$. Moreover,

$$\{Q, b(v)\} = T(v),$$

because the boundary of the cylinder $\exp(tv)$ viewed as a one-chain is equal to the same cylinder viewed as a point minus the trivial zero-width cylinder.

String theories are also referred to as topological, because of the following fact.

THEOREM 3. *The cohomology of the state space $H$ of a string background with respect to the differential $Q$ forms a TFT. Thus, the cohomology of $H$ has a natural structure of a Frobenius algebra.*

*Proof.* Two Riemann surfaces $\Sigma_1$ and $\Sigma_2$ which are diffeomorphic can be connected by a smooth path $C$ in the moduli space. Hence, for the corresponding states we have

$$|\Sigma_2\rangle - |\Sigma_1\rangle = |\partial C\rangle = Q|C\rangle,$$

which means that their $Q$-cohomology classes are equal. □

4.2. HIGHER BRACKETS

The state $|C\rangle$ is an operator from $H^m$ to $H^n$, which for $n = 1$ may be thought of as an $m$-ary operation on the space $H$. By the factorization axiom, the operation of sewing of chains $C$ in the moduli spaces corresponds to compositions of the corresponding operations on the space $H$. Respectively, any relation (involving compositions and boundaries) between chains in the moduli spaces produces an identity (involving compositions and the differential $Q$) for the corresponding operations on $H$. At the tree level, when we consider Riemann surfaces of genus 0 only, this algebraic structure on $H$ is rather tamable. This is because the topology of the finite dimensional moduli spaces $\mathcal{M}_{0,m+1}$ of isomorphism classes of $m+1$ punctured Riemann spheres takes over the situation.

Consider the following brackets:

$$[x_1, \ldots, x_m] = \text{``}|\mathcal{M}_{0,m+1}\rangle\text{''}(x_1, \ldots x_m), \qquad m \geq 2, \tag{6}$$

where $x_1, \ldots, x_m \in H$ are substituted on the right-hand side as arguments of the $\mathrm{Hom}(H^m, H)$, where the state $|\mathcal{M}_{0,m+1}\rangle$ lives. The quotes are due to the fact that the space $\mathcal{M}_{0,m+1}$ is not really a chain in $\mathcal{P}_{m+1}$: there is no natural mappings from $\mathcal{M}_{0,m+1}$ to $\mathcal{P}_{m+1}$. A standard escape is to impose these mappings as extra part of data. To preserve nice properties, this is achieved in the following two steps.

**Step 1**. Push the correspondence $C_\bullet \mathcal{P}_{m+1} \to \mathrm{Hom}(H^m, H)$ down to a mapping $C_\bullet \mathcal{P}'_{m+1} \to \mathrm{Hom}((H^{\mathrm{rel}})^m, H^{\mathrm{rel}})$ from chains on the quotient space $\mathcal{P}'_{m+1}$ of $\mathcal{P}_{m+1}$ by rigid rotations of the holomorphic disks to the space of multilinear operators on $H^{\mathrm{rel}}$. The latter is the subspace $H^{\mathrm{rel}}$ of vectors in $H$ which are *rotation-invariant*, i.e., stable under the operators $\exp(tT(v))$ of (4) and annihilated by the operators $B(tv)$ of (5) corresponding to rigid rotations $v \in S^1$. The pushdown is performed by pulling a chain $C$ in $\mathcal{P}'_{m+1}$ back to a chain $\widetilde{C}$ of the same dimension in $\mathcal{P}_{m+1}$, restricting the operator $|\widetilde{C}\rangle$ to $(H^{\mathrm{rel}})^m$ and projecting the value of the operator $|\widetilde{C}\rangle$ onto $H^{\mathrm{rel}}$ via the mapping $h \mapsto b(\partial/\partial\theta)h_0$, where $\theta$ is the phase parameter on the circle $S^1$ and $h_0$ is the rotation-invariant part of $h$ (which exists provided the action of $S^1$ on $H$ is diagonalizable).

**Step 2**. Map the finite dimensional moduli spaces $\mathcal{M}_{0,m+1}$ to the infinite dimensional quotient spaces $\mathcal{P}'_{m+1}$, so that gluing Riemann spheres in $\mathcal{M}_{0,m+1}$'s at punctures corresponds to sewing of Riemann spheres in $\mathcal{P}'_{m+1}$'s. Sewing in $\mathcal{P}'_{m+1}$'s can only be performed provided at least relative phases at sewn disks are given. The corresponding gluing operation should also be of this kind. Thus, the gluing operation takes us actually out of the spaces $\mathcal{M}_{0,m+1}$ to certain real compactifications of them, [5]. Such mappings $\mathcal{M}_{0,m+1} \to \mathcal{P}'_{m+1}$ exist. Zwiebach's string vertices [11] make up an example of those. Here we thereby allow certain freedom of their choice.

These additional data have been called a *closed string-field theory* in [5] after Zwiebach gave this title to the choice of his string vertices. After these modifications, we obtain brackets $[x_1, \ldots, x_m]$ defined on $H^{\mathrm{rel}}$.

THEOREM 4. *These brackets define the structure of a homotopy Lie algebra (see next section) on the space $H^{\mathrm{rel}}$.*

This result was obtained by Zwiebach in [11]. A mathematically rigorous proof of this theorem with the use of operads was given in [5]. This algebraic structure generalizes the trivial one of Corollary 2 to the case of string theory.

4.3. HOMOTOPY LIE ALGEBRAS

A *homotopy Lie algebra* is a graded vector space $H$, together with a differential $Q$, $Q^2 = 0$, of degree 1 and multilinear graded commutative brackets $[x_1, \ldots, x_m]$ of degree $3 - 2m$ for $m \geq 2$ and $x_1, \ldots, x_m \in H$, satisfying the

identities

$$Q[x_1, \ldots, x_m] + \sum_{i=1}^{m} \epsilon(i)[x_1, \ldots, Qx_i, \ldots, x_m]$$
$$= \sum_{\substack{k+l=m+1 \\ k,l \geq 2}} \sum_{\substack{\text{unshuffles } \sigma: \\ \{1,2,\ldots,m\} = I_1 \cup I_2, \\ I_1 = \{i_1, \ldots, i_k\}, \\ I_2 = \{j_1, \ldots, j_{l-1}\}}} \epsilon(\sigma)[[x_{i_1}, \ldots, x_{i_k}], x_{j_1}, \ldots, x_{j_{l-1}}],$$

where $\epsilon(i) = (-1)^{|x_1|+\cdots+|x_{i-1}|}$ is the sign picked up by taking $Q$ through $x_1$, $\ldots, x_{i-1}$, $|x|$ denoting the degree of $x \in H$, $\epsilon(\sigma)$ is the sign picked up by the elements $x_i$ passing through the $x_j$'s during the unshuffle of $x_1, \ldots, x_m$, as usual in graded algebra.

Note that for m=2, we have

$$Q[x_1, x_2, x_3] + (\pm[Qx_1, x_2, x_3] \pm [x_1, Qx_2, x_3] \pm [x_1, x_2, Qx_3])$$
$$= [[x_1, x_2], x_3] \pm [[x_1, x_3], x_2] \pm [[x_2, x_3], x_1],$$

which means that the graded Jacobi identity is satisfied up to a null-homotopy, the $Q$-exact term on the left-hand side.

### 4.4. HOLOMORPHIC STRING BACKGROUNDS AND HOMOTOPY COMMUTATIVE ALGEBRAS

According to general ideology, cf. Kontsevich [7] and Ginzburg-Kapranov [4], there are three principal types of homotopy algebras: homotopy Lie, homotopy commutative and homotopy associative. The first two types are dual in certain sense, the third one is self-dual. It is remarkable that this duality is implemented in string theory by passing from left-right movers' case to the chiral case, i.e., roughly speaking, from smooth "operator-valued differential forms" $|C\rangle$ on the moduli spaces to holomorphic ones.

More precisely, recall that the operators $T(v)$ and $b(v)$ of (4) and (5) are defined for tangent vectors $v$ to the Virasoro semigroup. Let us extend those operators to the complex tangent space to the Virasoro semigroup by $\mathbb{C}$-linearity. Since the Virasoro semigroup is a complex manifold, the complex tangent space splits into the holomorphic and antiholomorphic parts Vir and $\overline{\text{Vir}}$. Call a string background *chiral*, if $T(v) = 0$ and $b(v) = 0$ for $v \in \overline{\text{Vir}}$. Note that the first equation implies that the correspondence $C \mapsto |C\rangle$ determines a holomorphic mapping $C_\bullet \mathcal{P}_{m+n} \to \text{Hom}(H^m, H^n)$.

If we consider $m-2$-cycles (i.e., half-dimensional cycles relative to the boundary) in $\mathcal{M}_{0,m+1}$ instead of the fundamental cycle to define $m$-ary products as in (6), the operad approach of [5] will lead to the structure of a homotopy commutative algebra [3, 4]. This is the matter of the forthcoming paper [6].

TOPOLOGICAL FIELD THEORIES, STRING BACKGROUNDS, ETC.    11## 5. Topological Gravity and Gravity Algebras

A *topological gravity* is the same as a string background, except that it is based on a graded vector space $V$ which is not required to have a differential $Q$ and that the correspondence $C_\bullet \mathcal{P}_{m+n} \to \text{Hom}(H^{\otimes m}, H^{\otimes n})$ is replaced with

$$H_\bullet \overline{\mathcal{M}}_{m+n} \to \text{Hom}(V^{\otimes m}, V^{\otimes n}),$$

where $\overline{\mathcal{M}}_{m+n}$ is the Deligne-Knudsen-Mumford compactification of the moduli space and $H_\bullet$ stands for homology. Sewing in the factorization property should be replaced with gluing at double points similar to Step 2 in Section 4.2, but with no relative phases. This notion of a topological gravity is essentially the same as the notion of a homotopical field theory of Morava [8].

Another notion closely related to topological gravity is in certain sense a *dual topological gravity*, where the real compactification of [5] replaces the Deligne-Knudsen-Mumford one. At the tree level, when the Riemann surfaces have genus 0 and $n = 1$, this theory is Koszul dual to the one above in the sense of operads, see Getzler and Jones [3].

THEOREM 5. 1. *A string background based on a state space $H$ yields the structure of a dual topological gravity on the space $V$ of $Q$-cohomology of $H^{\text{rel}}$.*
2. *A dual topological gravity implies a TFT on its state space $V$.*

*Proof.* 1 becomes evident if we observe that the quotient of $\mathcal{P}_{m+n}$ modulo rigid rotations at each puncture is homotopically equivalent, respecting sewing, to (the real compactification of) $\mathcal{M}_{m+n}$.
2. A TFT is obtained by restricting a dual toplogical gravity to a correspondence $H_0(\mathcal{M}_{m+n}) \to \text{Hom}(V^m, V^n)$. □

At the tree level, a dual topological gravity also gives rise to a remarkable algebraic structure on $V$. This structure is called a *gravity algebra* and was introduced by Getzler [2]. It consists of an infinite number of multilinear brackets, staisfying quadratic equations. It would be interesting to describe the algebraic structure corresponding to a topological gravity, i.e., the structure of an algebra over the operad $H_\bullet \overline{\mathcal{M}}_{m+1}$, in similar terms.

### Acknowledgements

I am grateful to Y.-Z. Huang, T. Kimura, A. S. Schwarz, G. Segal and J. Stasheff for valuable discussions. I would like to thank the organizing committee of the DGM Conference in Ixtapa and Professors J. Keller and Z. Oziewicz, in particular, for their hospitality.